\documentclass[12pt,preprint]{aastex}

\shorttitle{Sideways displacement of penumbral fibrils}

\shortauthors{Jun Zhang, Leping Li and S. K. Solanki}

\begin{document}

\title{Sideways displacement of penumbral fibrils by the solar flare
on 2006 December 13}

\author{Jun Zhang\altaffilmark{1}, Leping Li\altaffilmark{1} and S. K. Solanki\altaffilmark{2}}

\altaffiltext{1}{Key Laboratory of Solar Activity, National
Astronomical Observatories, Chinese Academy of Sciences, Beijing
100012, China; E-mail: zjun@nao.cas.cn;lepingli@nao.cas.cn}
\altaffiltext{2}{Max-Planck-Institut f\"{u}r Sonnensystemforschung,
D-37191, Katlenburg-Lindau, Germany; E-mail: solanki@mps.mpg.de}

\begin{abstract}

Flares are known to restructure the magnetic field in the corona and
to accelerate the gas between the field lines, but their effect on
the photosphere is less well studied. New data of the Solar Optical
Telescope (SOT) onboard Hinode provide unprecedented opportunity to
uncover the photospheric effect of a solar flare, which associates
with an active region NOAA AR 10930 on 2006 December 13. We find a
clear lateral displacement of sunspot penumbral regions scanned by
two flare ribbons. In the impulsive phase of the flare, the flare
ribbons scan the sunspot at a speed of around 18 km s$^{-1}$,
derived from Ca II and G-band images. We find instantaneous
horizontal shear of penumbral fibrils, with initial velocities of
about 1.6 km s$^{-1}$, produced when a flare ribbon passes over
them. This velocity decreases rapidly at first, then gradually
decays, so that about one hour later, the fibrils return to a new
equilibrium. During the one hour interval, the total displacement of
these fibrils is around 2.0 Mm, with an average shear velocity of
0.55 km s$^{-1}$. This lateral motion of the penumbral fibrils
indicates that the magnetic footpoints of these field lines being
rearranged in the corona also move.

\end{abstract}


\section{INTRODUCTION}

Solar flares are one of the most spectacular phenomena in solar
physics, and quickly release magnetic energy stored in the corona in
a short time. Flare-associated changes in the magnetic field are
fundamental, and of importance in the general physics of magnetic
energy storage and explosive release. These changes in photospheric
magnetic fields largely include two topics: long-term changes before
the flare and short-term changes during the flare, among which the
short-term changes mean magnetic changes that occur on the flare
time-scale \citep{zhao09}. The rapid short-term magnetic field
changes in the course of flares was first reported by
\citet{patt81}. However, \citet{patt84} later recognized that the
reported magnetic transients were not real changes in the magnetic
field but could be explained by transient emission of the Fe I 5324
\AA~line which was used to produce the magnetograms. \citet{qiu03}
provided a scenario that the observed transient polarity reversal in
MDI magnetograms is likely to be produced by distorted measurements
when the Ni I 6768 \AA~line comes into emission or strong central
reversal as a result of non-thermal beam impact on the atmosphere in
regions of strong magnetic fields. They called this transient
polarity reversal as magnetic ``anomaly".

On the other hand, \citet{koso99} detected rapid variations of the
photospheric magnetic field during an X-class flare in AR 8210 on
1998 May 2, and \citet{came99} found a significant change in the
longitudinal magnetic fields in NOAA AR 6063 during an X9.3 flare.
\citet{sudo05} further presented new evidence of longitudinal
magnetic field changes accompanying 15 X-class flares. Associated
with Bastille Day flare/CME event \citep{zhang01}, \citet{koso01}
detected two types of rapid magnetic changes : irreversible changes
and magnetic transients. Recently, \citet{wang04} and \citet{deng05}
found penumbral decay and neighboring umbral core darkening rapidly
right after the flares. \citet{liu05} further studied this type of
short-term evolution of sunspot structure associated with seven
major flares, and proposed accordingly a reconnection picture to
interpret the findings. Moreover, \citet{chen07} found that the
rapid and permanent structural changes are evidenced in the time
profile of white light mean intensity and are not likely resulted
from flare emission. In addition, \citet{wang09} and \citet{zhao09}
discovered rapid, significant and persistent changes and flare
induced signals in polarization measurements during flare events.

Here, we report on observations from Hinode spacecraft of sideways
displacement of penumbral fibrils driven by a solar flare on Dec.
13, 2006. This displacement carries potentially interesting
information about the flare induced variations of the magnetic
fields and sunspots.

\section{OBSERVATIONS AND ANALYSIS}

Active region NOAA AR 10930 displayed the evolution of a complex
magnetic structure, surrounded by many small dark pores and bright
magnetic faculae \citep{zhang07,li09}. Images of the X3.4 flare
related to the active region on December 13, 2006, were obtained by
the Solar Optical Telescope \citep[SOT,][]{tsun08} onboard Hinode in
two spectral filters: the molecular line G-band and Ca II H. In each
filter the images were obtained every 2 minutes, and the pixel size
of the 2048${\times}$1024 images was 0.11 arcsec.

A first signature of the flare appeared in the Ca II H images at
about 2:08 UT between two sunspots with the opposite magnetic
polarity \citep{koso07}. The energy release was probably triggered
by strong shearing flows around the following sunspot
\citep{zhang07}. The flare emission quickly extended during the next
10 min, forming a two-ribbon structure. Figure 1 displays a G-band
(top) and a co-temporal Ca II image (bottom) from {\it Hinode}/SOT.
The three ellipses `E1', `E2' and `E3' outline the flare-affected
regions determining from time slice maps, and overplotted on the Ca
II image. The dash-dotted curves represent the leading edge of the
two ribbons at different times. At 02:22 UT, the two flare ribbons
(R1 and R2) first became visible in the G-band images, indicating
that the perturbation extends into the photosphere. R1 moved
laterally across the penumbra of the sunspot P1 with a velocity of
18 km $^{-1}$ in the first 2 minutes, after which its horizontal
propagation velocity decreased rapidly. The ribbons on the Ca II
images first appeared at 02:16 UT, 6 minutes earlier than on the
G-band. From 02:16 to 02:20 UT, the apparent motion velocity of R1
was about 12 km s$^{-1}$, and from 02:20 to 02:24 UT, R1 scanned
across the penumbra with a velocity of 19 km s$^{-1}$. Then the
chromospheric ribbon also quickly slowed down. In the course of the
scanning, the two ribbons expanded in length and separated from each
other, subsequently they intrude from the penumbrae into the umbrae
of the sunspots. The dotted and the dashed lines labeled `a' to `g'
and `ref1', `ref2', cut the penumbrae fibrils, respectively. The
lines `a' to `g' lie in the parts of the penumbra where we see an
immediate influence of the flare during the passage of the ribbons
(see the accompanying movie), while `ref1' and `ref2' are located in
parts of the penumbra that are not crossed by the flare ribbons.
Time slices of the measured quantities along these lines are shown
in Figs. 2-4.

We display four time-slice maps taken from the G-band images in Fig.
2. From top to bottom, these maps correspond to the four lines
`ref1', `a', `b' and `c', respectively. The X-axis is the distance
($\sim$22 Mm) of the lines from left to right, which cut the
penumbra of the largest sunspot `P1'. The Y-axis of each map denotes
time, running from 2006 Dec. 13, 02:00 to 04:00 UT (from bottom to
top). Bright horizontal strips on maps (a)-(c) represent the flare
ribbon `R1' at the time when it crosses that particular cut, while
moving from the neutral area of the active region to the umbra of
the largest sunspot `P1'. The top map indicates the G-band signals
at the position `ref1' where it is not disturbed by the flare ribbon
R1, and the others maps (a)-(c), the signals affected by the ribbon.
From map (ref1), we find no shear motion of the penumbral fibrils.
However, from maps (a)-(c), we find that there is a rightwards tilt
of these fibrils before the arriving of the flare ribbon R1 (see the
blue and black arrows in Fig. 2b). The inward motion of bright/dark
structures can explain the rightwards tilt in the time-slice maps
(a), (b), (c) between 02:00 UT and 02:14 UT (e.g. see the blue
arrow). However, the rightwards tilt between 02:14 UT and 02:22 UT
(see the black arrow) is due to lateral rightwards displacement of
the penumbral fibrils and it is related to the flare. Moreover, we
find distinct opposite sideward displacement (see the white arrows)
when the flare ribbon first scans the fibril in the photospheric
layers. The whole fibril is offset perpendicular to its long
direction and not just the part at which the flare ribbon currently
resides (see the accompanying movie). Numbers 1-6 mark the positions
where the lateral velocities are plotted in the bottom of Fig. 5.
The lateral motion lasts for about one hour. This is much longer
than the duration of the flare ribbon resides in the penumbra, which
is 4-10 minutes, depending on which part of the penumbra we are
considering. Although the signals on numbers `1' and `2' still
display shear motion one hour after the initial shear, this motion
is mainly contributed from the outward movement of bright material
along the penumbral fibrils, seen from G-band movie. In the one hour
interval from the original shear, the displacements at the six
positions 1-6 are 2.1 Mm, 2.1 Mm, 2.0 Mm, 1.9 Mm, 1.8 Mm and 1.4 Mm,
respectively. Evidently, the displacement of the penumbral fibrils
started immediately after the flare ribbon reached the region.

Figure 3 shows time-slice maps along lines `d' and `e' in the
penumbra of the opposite polarity spot during and after the
passage of the flare ribbon R2. This flare ribbon looks like a
bright horizontal strip on each map, at the time when it crosses
that particular cut, while moving from the central area of the
active region to the umbra of the smaller sunspot `P2'. Figure
3(d) shows that the fibrils along this cut display a lateral
motion. This motion lasts for about 50 minutes with a final
displacement of 1.6 Mm. Figure 3(e) shows the displacement of a
lightbridge which separates the two sunspots (P2 and P3). The
lightbridge displays a smaller displacement (0.7 Mm) in the 50
minutes interval.

Figure 4 shows three time-slice maps (corresponding to the three
lines `ref2', `f' and `g' in Fig. 1) taken from the G-band images.
The top panel of Fig. 4(ref2) confirms that the penumbral fibrils in
regions unaffected by the flare ribbon remain at the line `ref2'
position, quite stable in the two hour interval. The penumbral
fibrils cut by the lines `f' and `g' are parallel to the umbral
boundary of sunspot P2. When flare ribbon R2 scans the penumbra, the
fibrils displayed in map `f' shear with a velocity of 1.1 km
s$^{-1}$ in the first 10 minutes. For map `g', the shear velocity of
the penumbral fibrils is 0.8 km s$^{-1}$. The lateral motion last
for about 30 minutes, along both cuts, which is shorter than that of
the fibrils displayed in Figs. 2 and 3. The top panel of Fig. 5
displays soft X-ray flux (solid curve) of the flare measured by
GOES-12 and lateral horizontal speed (dotted curve) at the position
of number `3' denoted in Fig. 2. Light gray area represents the
period at which the flare is strongest. The shear appears suddenly
at 02:24 UT, with a peak velocity of 1.6 km s$^{-1}$, then the
velocity decreases rapidly, and lasts for about one hour. On the
bottom panel, the filled circles represent the peak lateral
velocities between 2:24 and 2:28 UT at the six positions showed in
Fig. 2, and the empty circles, the mean lateral velocities between
2:24 and 3:24 UT. The peak velocities of all the six positions are
around 1.6 km s$^{-1}$, while the mean velocity, 0.55 km s$^{-1}$.

\section{DISCUSSION AND CONCLUSIONS}

The observations of the solar flare of December 13, 2006 from Hinode
reveal a flare-induced displacement of penumbral fibrils, as seen in
series of G-band images. The displament was started locally
immediately after the passage of a flare ribbon. Fibrils not crossed
by the flare ribbon are not seen to move. The passage of the flare
ribbon produces a strong acceleration of the fibrils transverse to
their elongated direction almost instantaneously. This results in a
peak velocity reaching up to 1.6 km s$^{-1}$. The maximum value of
the horizontal displacement reaches 2.0 Mm, although at most
locations the displacement is smaller, so that it is perhaps not
surprising that the discovery of such photospheric effects of flares
had to await the advent of data with consistent high resolution
approaching 0.2$''$ as provided by Hinode.

\citet{wang04} studied three X-class flares in the active region
NOAA 10486 and found that the penumbral segments decayed rapidly
right after the flares; meanwhile, the adjacent umbral cores became
darker. The location of such a disappearance of penumbral structure
coincides with, or is close to, the flare kernel. Moreover, these
changes are permanent, e.g., they remained well after the flares and
did not go back to the preflare level. They put forward the
interpretation of the results: magnetic fields become more vertical
after flares, which cause penumbrae to decay. The umbrae become
darker as a result of enhancement of longitudinal magnetic field.
However, in this study, we find the sideways displacement of the
penumbral fibrils associated with the flare instead of the penumbral
decay observed by \citet{wang04}.

The structure of the magnetic field in the penumbra is complex, with
interlaced fibrils of different magnetic inclination
\citep[e.g.][]{lites93,title93,sola93,sol03}, which are related to
fibril brightness in inner and lateral penumbra. As the cause of the
shift, we propose the motion of field lines rearranged by the
flares. \citet{zirin84} have provided evidence that the coronal
magnetic field rearranges itself significantly in the course of a
flare. Such a rearrangement also leads to force that attempt to move
the photospheric footpoints of the affected fields lines. Since the
gas density is 8 orders of magnitude higher in the photosphere than
in the corona and the pressure is 10$^{5}$ times higher, it is
generally assumed that the field in the photosphere cannot be moved
(line-tying). In a sunspot, however, the plasma \textbf{${\beta}=1$}
\citep{solan93}, so that magnetic field can also move the plasma
bodily.

Our results therefore test the line-tying condition often involved
in coronal magnetic field studies \citep{hood86,zwei92} and shows
that although linetying is not exactly fulfilled, even in the case
of an X-class flare and the field lines anchored in the penumbra,
their footpoints move by less than 2 Mm. This suggests that the
line-tying condition is relatively well fulfilled, e.g. for
reconnection work. Furthermore, the sideways displacement occurred
simultaneous with the passage of the flare ribbon may be the result
of the lower reconnection site during the flare event.

In addition, under some assumptions, we estimate the value of the
magnetic energy needed to result in the sideways displacement of the
penumbral fibrils. Here, we use 10$^{8}$ km$^{2}$ to be the area of
these penumbral fibrils, and 10$^{3}$ km, the height. As the density
of the photosphere is about 10$^{-9}$ g cm$^{-2}$, then the mass of
these penumbral fibrils is 10$^{17}$ g. The maximum velocity of
these penumbral fibrils is 1.6 km s$^{-1}$. Therefore, the kinetic
energy of these penumbral fibrils are about 10$^{27}$ erg. That is
to say, more than 10$^{27}$ erg  magnetic energy will lead to the
sideways displacement of the penumbral fibrils. Actually, the
magnetic energy of lots of big flares are beyond this value. With
higher spatial resolution observations (such as Hinode), we may find
more flare events with sideways displacement of the penumbral
fibrils. More major flare events need to be examined if the revealed
magnetic changes in the course of flare by this study are common.

\acknowledgments

The authors are indebted to the {\it Hinode} teams for providing the
data. {\it Hinode} is a Japanese mission developed and launched by
ISAS/JAXA, with NAOJ as domestic partner and NASA and STFC (UK) as
international partners. It is operated by these agencies in
co-operation with ESA and NSC (Norway). J. Z. is supported by the
National Natural Science Foundations of China (G40809161), the CAS
Project KJCX2-YW-T04, and the National Basic Research Program of
China under grant G2006CB806303.

\clearpage

\begin{figure}
\epsscale{0.50} \plotone{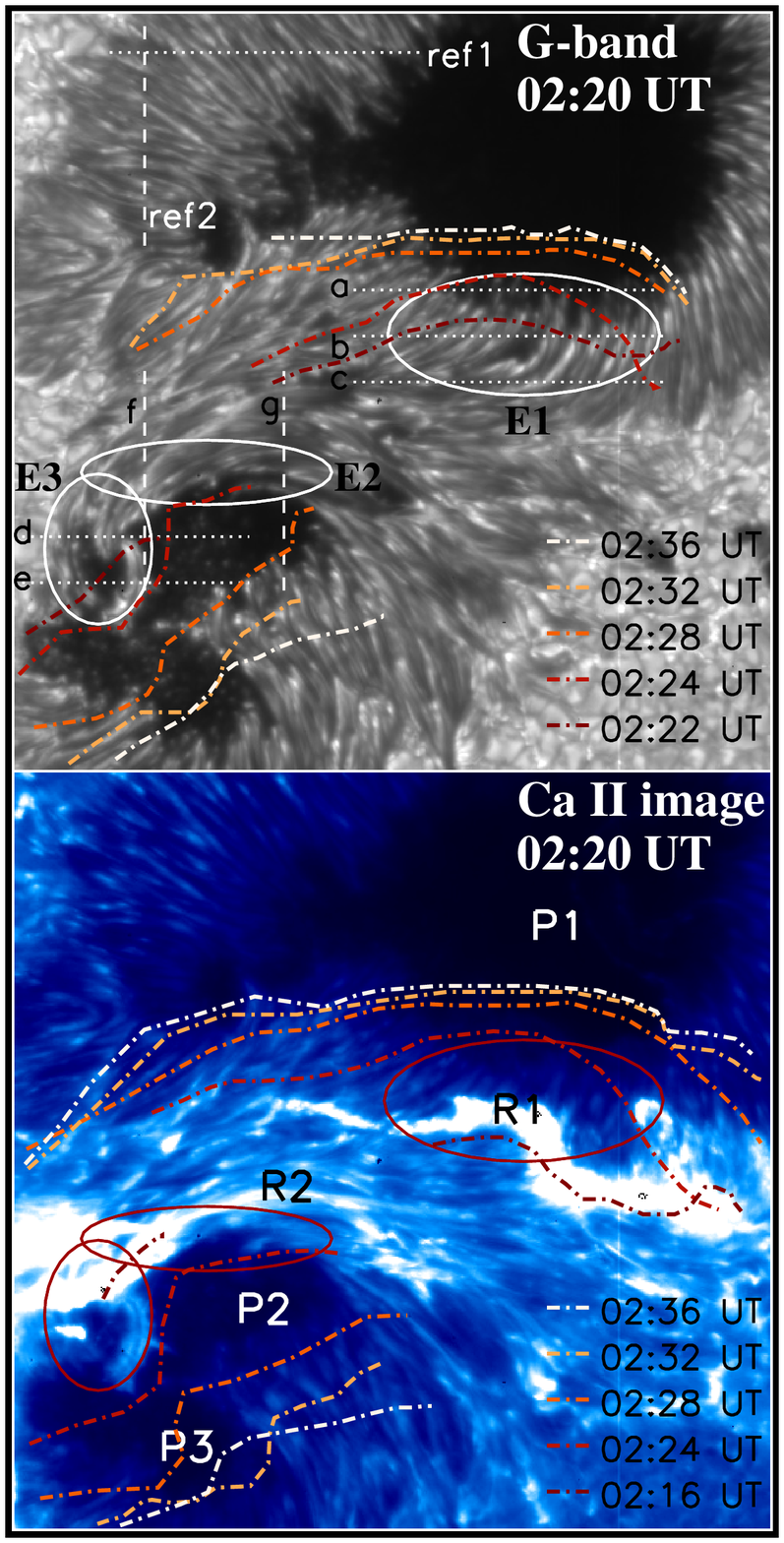} \caption{A G-band (top) and a
co-temporal Ca II image (bottom) from {\it Hinode}/SOT showing the
penumbral structures of sunspots (`P1', `P2' and `P3' in the Ca II
image) belonging to the active region NOAA AR 10930 on 2006 Dec. 13.
The dash-dotted curves represent the leading edges of the two
ribbons (marked by `R1' and `R2' in the Ca II image) while they scan
these penumbral fibrils (see also the accompanying movie), with
different colors indicating the different time. The dotted and the
dashed lines cut the penumbral fibrils. Time slices of measured
quantities along these lines are shown in Figs. 2-4. The three
ellipses `E1', `E2' and `E3' on the G-band outline flare-affected
regions determining from the time slice maps, and are overplotted on
the Ca II image. The field-of-view of both frames is
55$''$${\times}$55$''$. \label{fig1}}
\end{figure}

\begin{figure}
\epsscale{0.8} \plotone{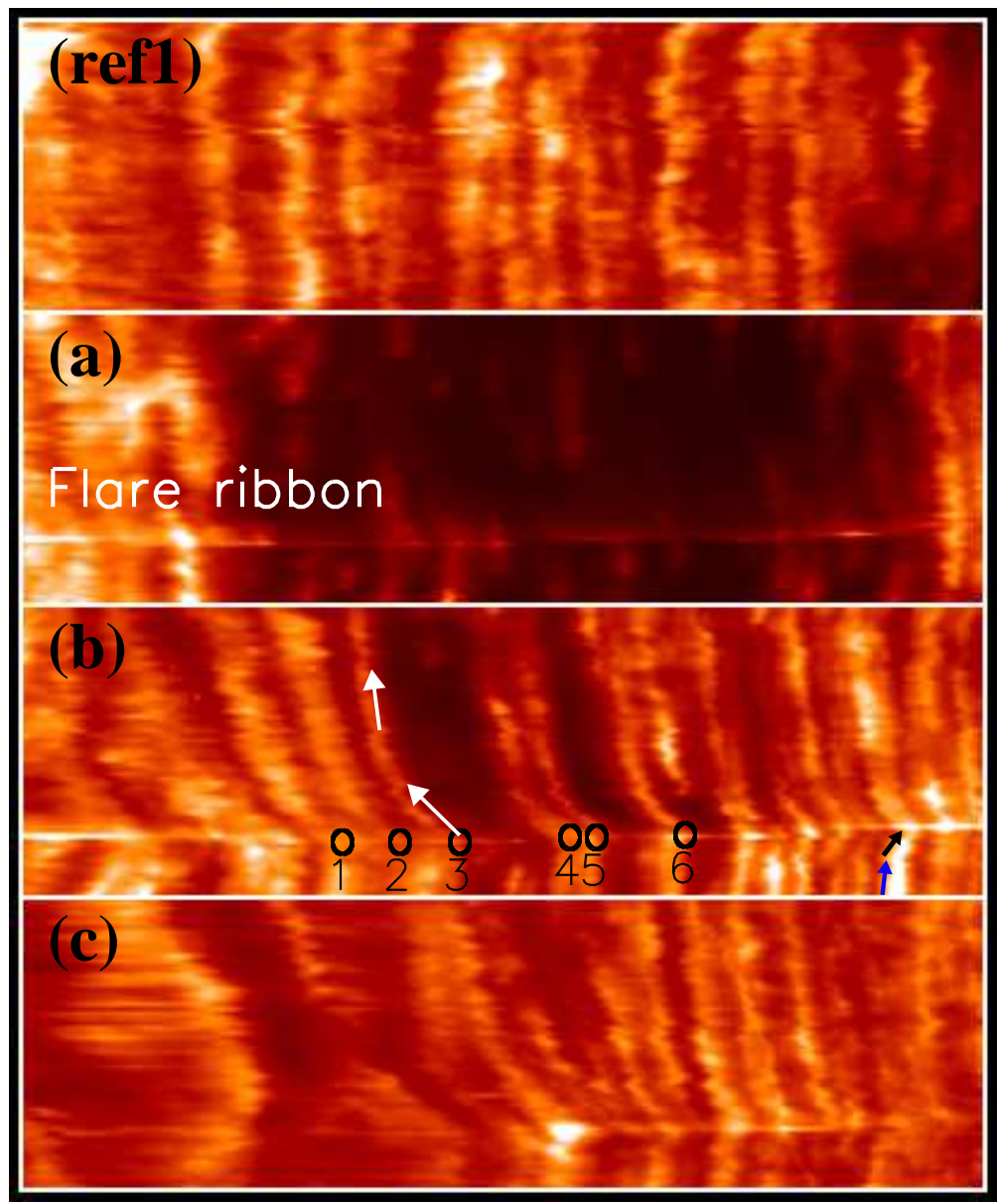}
\caption{{\it From top to bottom}: four time-slice maps, which
correspond to the four lines `ref1', `a', `b' and `c' (see Fig. 1),
respectively, taken from the G-band images. The X-axis is the
distance ($\sim$22 Mm) of the lines from left to right. Line `ref1'
is given for a reference. Lines `a', `b' and `c' cut the penumbra of
the largest sunspot `P1'. The Y-axis of each map denotes time,
running from 2006 Dec. 13, 02:00 to 04:00 UT. Circles numbered by
1-6 show the position of the shear velocities plotted in the bottom
of Fig. 5. Bright strips on maps `a-c' represent the flare ribbon
`R1' while it moves from the central area of the active region to
the umbra of the largest sunspot `P1'. The slopes of the arrows
indicate the speed of motion in the direction of the cut.
\label{fig2}}
\end{figure}


\begin{figure}
\epsscale{0.8} \plotone{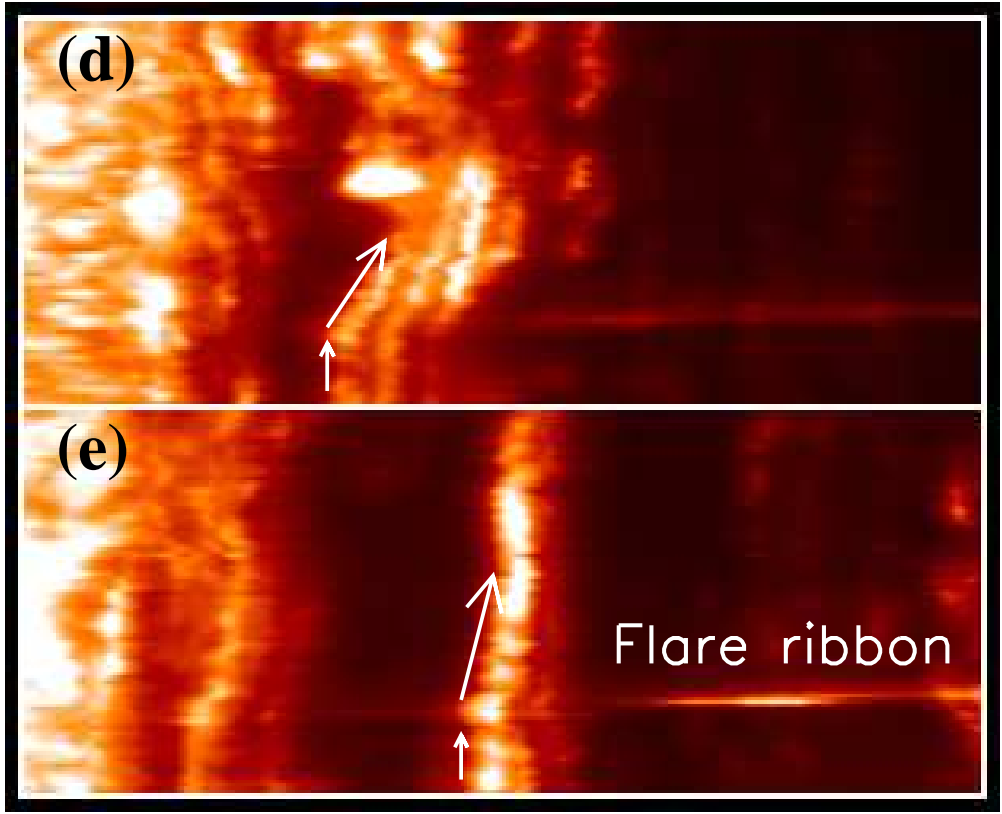}
\caption{Similar to Fig. 2 but for the two horizontal lines `d' and
`e' (see Fig. 1), with X-axis representing the distance ($\sim$16
Mm) from left to right. These lines cut the penumbra of the smaller
sunspot `P2' from the outer to the inner boundary. The Y-axis of
each map denotes time, running from 2006 Dec. 13, 02:00 to 04:00 UT
(from bottom to top). A bright strip on each map represents the
flare ribbon `R2' while it moves from the central area of the active
region to the outer boundary of the smaller sunspot `P2'. The slopes
of the arrows indicate the speed of motion in the direction of the
cut. \label{fig3}}
\end{figure}

\begin{figure}
\epsscale{0.8} \plotone{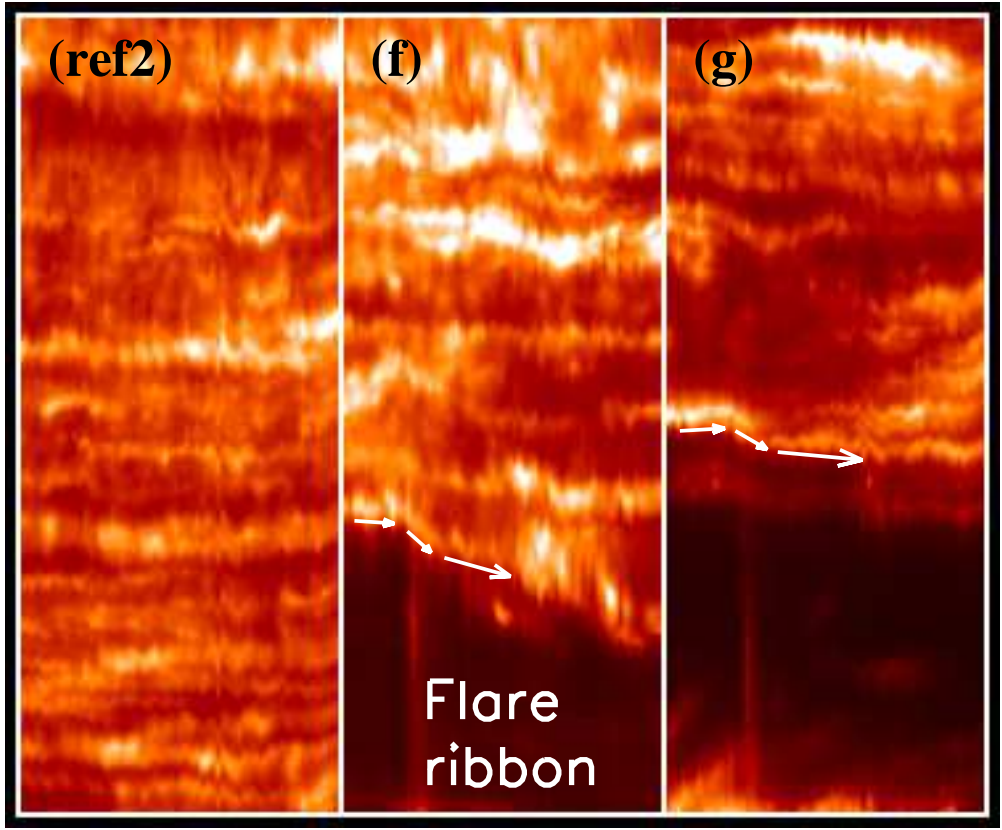}
\caption{Similar to Fig. 2 but for the three vertical lines `ref2',
`f' and `g' (see Fig. 1). The X-axis of each map denotes time,
running from 2006 Dec. 13, 02:00 to 04:00 UT (from left to right).
The Y-axis represents the distance ($\sim$16 Mm) of the lines. Line
`ref2' is considered for a reference. Lines `f' and `g' cut the
penumbra of the smaller sunspot `P2'. A bright strip on each map
represents the flare ribbon `R2'. The slopes of the arrows indicate
the speed of motion in the direction of the cut which are on the
umbral boundary of `P2'. \label{fig4}}
\end{figure}

\begin{figure}
\epsscale{0.8} \plotone{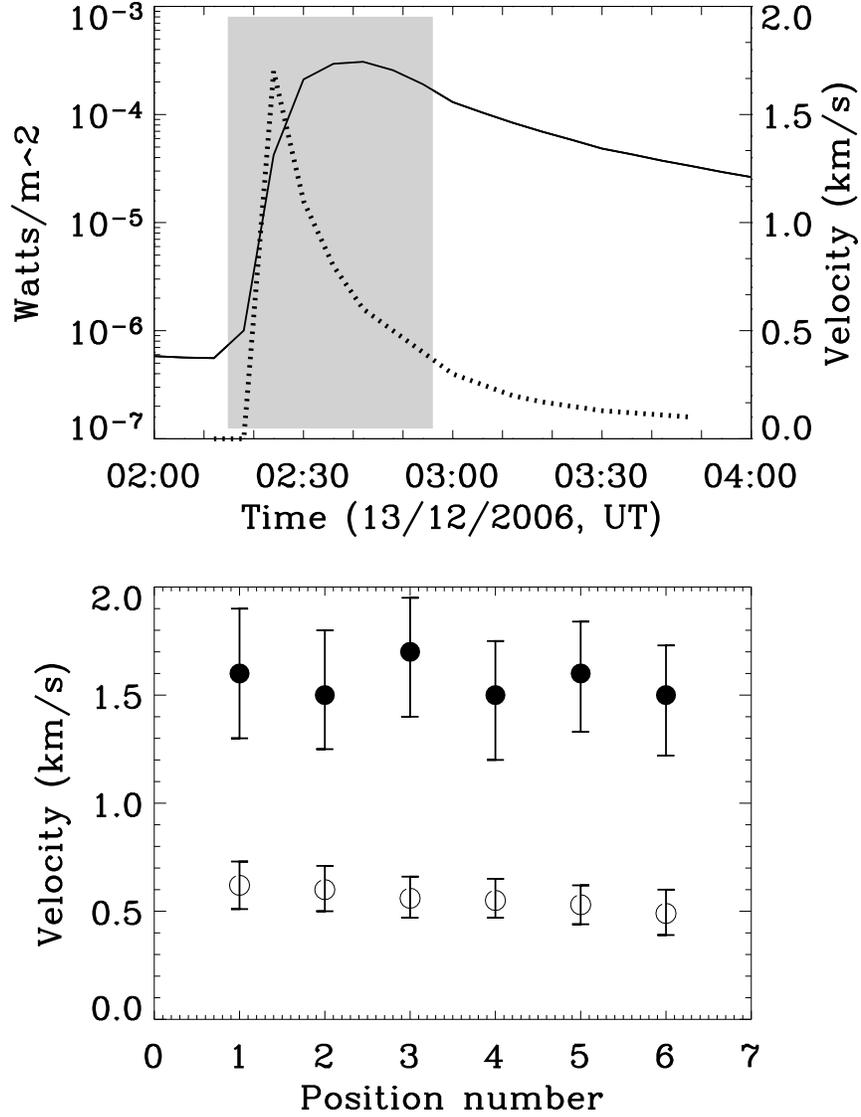} \caption{{\it Top}: Soft X-ray
flux (solid curve) measured by GOES-12 and shear velocities
(dotted curve) at the position of number `3' denoted in Fig. 2.
Light gray area represents the flaring activity period. {\it
Bottom}: Estimates of the peak shear velocities (solid circles)
between 2:24 and 2:28 UT, as well as mean shear velocities (hollow
circles) between 2:24 and 3:24 UT at the six positions (see Fig.
2) in the penumbra of the largest sunspot. \label{fig5}}
\end{figure}

\clearpage

\end{document}